\RequirePackage{fix-cm}
\documentclass[twocolumn]{svjour3}          
\smartqed  
\usepackage{graphicx}
\usepackage{url}
\usepackage{nameref}
\usepackage{multirow}
\usepackage{diagbox}
\usepackage{tikz}
\usetikzlibrary{fit,positioning,arrows,automata,calc,shapes,shapes.geometric}
\usetikzlibrary{decorations}
\usepackage[caption=false]{subfig}
\usepackage{makecell}%
\usepackage[final]{changes}
\newcommand{\human}[3][1]{
\node[scale=#1] (#2) at (#3) {
\begin{tikzpicture}
\node[circle,fill,minimum size=5mm] (head) {};
\node[rounded corners=2pt,minimum height=1.3cm,minimum width=0.4cm,fill,below = 1pt of head] (body) {};
\draw[line width=1mm,round cap-round cap] ([shift={(2pt,-1pt)}]body.north east) --++(-90:6mm);
\draw[line width=1mm,round cap-round cap] ([shift={(-2pt,-1pt)}]body.north west)--++(-90:6mm);
\draw[thick,white,-round cap] (body.south) --++(90:5.5mm);
\end{tikzpicture}};
}

\definechangesauthor[color=black!60!green]{aby}

\journalname{Draft}
\begin{document}

\title{On-the-fly Detection of User Engagement Decrease in Spontaneous Human-Robot Interaction using Recurrent and Deep Neural Networks 
}

\titlerunning{On-the-fly Detection of User Engagement Decrease in Spontaneous Human-Robot Interaction}        

\author{Atef~Ben-Youssef         \and
        Giovanna~Varni         \and
        Slim~Essid \and
        Chlo\'e~Clavel 
}


\institute{A. Ben-Youssef \at
              Telecom ParisTech, 75013 Paris, France \\
              \email{atef.benyoussef@telecom-paristech.fr}           
           \and
           G. Varni \at
              Telecom ParisTech, 75013 Paris, France \\
              \email{giovanna.varni@telecom-paristech.fr}            
            \and
           S. Essid \at
              Telecom ParisTech, 75013 Paris, France \\
              \email{slim.essid@telecom-paristech.fr}
            \and
           C. Clavel \at
              Telecom ParisTech, 75013 Paris, France \\
              \email{chloe.clavel@telecom-paristech.fr}     
}

\date{}

\maketitle

\begin{abstract}
In this paper we consider the detection of a decrease of engagement by users spontaneously interacting with a socially assistive robot in a public space. 
We first describe the UE-HRI dataset that collects spontaneous Human-Robot Interactions following the guidelines provided by the Affective Computing research community to collect data ``in-the-wild''.  
We then analyze the users' behaviors, focusing on proxemics, gaze, head motion, facial expressions and speech during interactions with the robot. 
Finally, we investigate the use of deep leaning techniques (Recurrent and Deep Neural Networks) to detect user engagement decrease in real-time. 
The results of this work highlight, in particular, the relevance of taking into account the temporal dynamics of a user's behavior. 
Allowing 1 to 2 seconds as buffer delay improves the performance of taking a decision on user engagement. 

\end{abstract}

\section{Introduction}
\label{intro}

Socially assistive robots (SAR) should be able to communicate and cooperate with humans in order to provide assistance, coaching, companionship, support for convalescence, rehabilitation, learning, or therapeutic aid, etc. (e.g.\cite{Feil-Seifer,Tapus2008}). 
SAR deployed in public spaces 
have considerable potential for providing the humans with whom they engage, with a multitude of services: welcoming them, giving them recommendations or interacting in a personalized way \cite{Giuliani2013,Foster2016,bohus2009open-world-dialog,Bohus:2009:MME}. 
These types of robot employ short-term adaptation in order to keep the user's attention 
and achieve their goal 
of assisting them through social interaction.
They are equipped with sensors combined with software modules to track humans and inform the interaction process. These modules can for instance track faces, recognize speech, and synthesize speech synchronized with animation. Extracting basic information such as facial expressions, gaze, and head motions allows the robots to better understand the person. 
Processing this information serves more sophisticated modules that analyze emotions, mood, affective state, and user's engagement in order to give appropriate responses. 

This study focuses on real-time detection of user's engagement decrease during a social interaction with a robot in a public space. 
In public space settings, it is not easy for the robot to achieve its goal in spontaneous social interaction, where participants are free to treat the robot as they like and leave the interaction when they wish \added[id=aby]{\cite{Ben-Youssef2019ieee}}. 
Recognizing user's engagement state represents a key issue in socially assistive robotics. 


For this study, we recorded a multimodal dataset of spontaneous interactions with the humanoid robot Pepper\footnote{https://www.softbankrobotics.com/emea/en/pepper} \cite{Ben-Youssef2017}. 
In keeping with the current emerging trend in Affective Computing, this dataset consists of data collected “in-the-wild” \cite{schuller2016multimodal}. 
It comprises 278 interactions where the users were free to participate in the interaction if they wished to and free to leave it when they wanted to, and where they were left to behave without unconstraints. 
Multimodal information describing the user's behavior (\emph{i.e.} distance to the robot, gaze and head motion as well as facial expressions and speech features) was thus synchronously recorded.
We analyze the dataset, focusing on the non-verbal behavior displayed by the users.
We then make use of data-driven methods for detecting engagement decrease. Such methods rely on a ground-truth obtained by manual annotation of the engagement. 
Perceived engagement can be a subjective observation. For this reason, each interaction was annotated independently by two annotators: a researcher who knew the purpose of the work and an uninformed one who did not. 

This paper is organized as follows. Section~\ref{sec:relatedwork} presents the related work on user engagement in Human-Robot interaction (HRI). 
Section~\ref{sec:SpontaneousHRI} describes the dataset of spontaneous HRI. 
Section~\ref{sec:analysis} focuses on the analysis of user engagement decrease.
Section~\ref{sec:modelling} describes our approach to detect the decrease of user engagement. 
A discussion is presented in Section~\ref{sec:discussion}. 
Finally, conclusions are drawn in Section~\ref{sec:conclusion}.

\section{Related Work}
\label{sec:relatedwork}

\subsection{Socially Assistive Robots in Public Space}

\replaced[id=aby]{SAR}{Socially assistive robots (SAR)} are robots providing assistance to human users through social interaction \cite{Tapus2008}. These robots are designed e assist users by creating effective interactions \cite{Feil-Seifer}.  
SAR deployed in real world settings need to secure and maintain the user’s engagement. 
Pitsch \emph{et al.} \cite{Pitsch2009} analyzed interactions between a robot deployed as a guide in a museum, and visitors. 
They found that the first five seconds of the interaction \replaced[id=aby]{ }{ have} had a relevant impact on the user's engagement during the interaction (\emph{e.g.} leaving/staying, responsiveness, exchanging rituals). 
Gehle \emph{et al.} \cite{Gehle2017} likewise analyzed the interaction opening strategies of a robot playing the role of a museum's guide, in its interaction with visitors. 
Hayashi \emph{et al.} \cite{Hayashi2007} proposed to use robots in train stations to assist passengers. Their goal was to identify the best way to provide users with travel information. They compared the use of one vs two robots.
The findings of this study showed that the most effective way of attracting people's interest was by presenting information using two humanoid robots rather than one. They reported also that the interactivity was useful in giving the feeling of talking with robots. 
Another interesting scenario is the use of SAR to provide shopping information to customers \cite{Foster2016,Kanda2009}.
The MuMMER project aims to develop a socially intelligent humanoid robot that is able to operate in a public shopping mall \cite{Foster2016}.
In \cite{Kanda2009}, SAR were designed to naturally interact with customers and to \deleted[id=aby]{affectively} provide shopping information. 
In public spaces\deleted[id=aby]{ under the hotel's context}, SAR could inspire the design of hotel-assistive robots \cite{Feil-Seifer}. 

In such application contexts, robots are expected to respond appropriately to the users' behavior and engage them in stimulating experiences \cite{Clavel2016,Dominey2008}. In particular, they should be able to monitor a user’s state of engagement in order to be able to react to possible signs of disengagement in such a way as to maintain their interest.
In real world settings, one of the challenges is to deal with the dynamic and flexible nature of human behavior in order to secure and maintain users' engagement in their interaction with SAR.

Tackling these challenges, our research aims to detect user engagement states in real-time in order to assist humans for the purpose of providing such public services. The proposed detection model integrates data on the temporal dynamics of engagement behavior, with the multimodal data collected \textit{in-the-wild}.

\subsection{Engagement and Disengagement in HRI}
\label{ssec:engagementHRI}

The engagement was defined in human-computer interaction by Sidner \cite{Sidner2005} as \textit{``the process by which individuals in an interaction start, maintain and end their perceived connection to one another''}; and by Poggi \cite{pog07} as \textit{``the value that a participant in an interaction attributes to the goal of being together with the other participant(s) and of continuing the interaction''}.
This concept of engagement has been explored from different perspectives with regard to humans interacting with social robots or virtual agents \cite{Corrigan2016}. More specifically, a focus has been put on user engagement prediction \cite{Foster2017,Castellano2012}, the analysis of the emergence of engagement \cite{Vaufreydaz2016,Pitsch2009}, the identification of the addressees to interact with \cite{Li2012}, and the study of the relationship between personality and engagement \cite{Celiktutan2017,Ivaldi2017} and engagement perception \cite{Hall2014}.
Similarly, disengagement has been tackled in many studies by analyzing 
interaction problems, the time of their occurrence and their causes \cite{Andrist2017,Trung2017}, 
the dynamics of affective states \cite{DMello2012,Bosch2015} and the prediction of disengagement \cite{Bohus2014,Leite2015}.
The most crucial causes of interaction problems are found to be the limitations of the systems used to detect social signals and of the interaction models. For example, it was reported that the most frequent causes were 
the engagement model, face tracker, turn-taking model, or
speech recognition issues, misunderstanding, 
lack of adaptation, repetition and long pauses, 
over-fragmentation, over-clarity, over-coordination, over-directedness, insufficient or exaggerated state-of-mind updates and repair requests \cite{Bohus2014,Andrist2017,Martinovski2003}. 

Humans behave differently during social norm violations and technical errors in HRI \cite{Trung2017}. It was shown that the automatic detection of these errors based on human behavior works to some extent. The performance of error detection is better when the robot knows the human with whom it is interacting 
. Detecting social norm violation is harder than detecting technical failures. We conclude from the work of Trung \textit{et al.} \cite{Trung2017} that detecting disengagement in social interaction with a robot is difficult.


The most common features used in these studies to asses engagement and disengagement were, among others, gaze \cite{Anzalone2015,Ivaldi2017,Rich2010,Sidner2005,Leite2015}, 
head motion \cite{Anzalone2015,Leite2015}, 
face \cite{Bohus2014,Leite2015,liu2018predicting}, 
posture \cite{Anzalone2015,Leite2015}, 
speech \cite{Ivaldi2017,Leite2015}, and distance  \cite{Vaufreydaz2016}.
Other, more subtle, features were also included: semantics, attention, emotions and affects \cite{Bohus2014,Leite2015,DMello2012,Bosch2015}.
\added[id=aby]{In a previous study, we show that the use of combined multimodal features effectively improves the performance of a user engagement breakdown system \cite{Ben-Youssef2019ieee}.}
Combining features from two or more modalities allows one to achieve better results in engagement detection/prediction, compared to the use of features from only one modality.
Kendon \cite{Kendon1967} analyzed gaze and speech. He found that the speakers looks at each other during fluent speech and at the end of sentences, but look away during hesitations or unfluent speech. This type of social signal is probably relevant information to evaluate the engagement level during the interaction. 
Prosody, articulation, voice-quality related features, linguistic analysis as well as facial expressions and gaze were used to detect interest in \cite{Schuller2007}. 


To model user engagement in HRI, researchers have considered a subset of systems going from rule-based to machine-learning-based. Machine learning approaches have been compared to rule-based approaches in \cite{Foster2017}.
It has been shown that the rule-based classifiers have a competitive performance compared to the set of supervised classifiers trained on a small labeled corpus. The authors found that Conditional Random Fields (CRF), which give an accuracy of 61.5\% and F1-score of 0.61, is a much more stable classifier than others. 
Machine learning approaches are the most commonly used for automatically predicting engagement in HRI.
By comparing logistic regression and boosted decision tree models in \cite{Bohus2014}, the logistic regression model was selected for managing disengagement decisions. 
In \cite{Bohus2009}, Bohus \textit{et al.} used a frame-by-frame binary classification scheme using a maximum entropy model to predict engagement intentions.
Leave-one-out cross-validation using Support Vector Machines (SVMs) was used in \cite{Castellano2012,Schuller2007,Leite2015}. 
SVMs with a polynomial kernel were successfully used to recognize \added[id=aby]{the} interest in \cite{Schuller2007}.
The problem to address the engagement of only one user or more than one in interaction was studied by Leite \textit{et al.} \cite{Leite2015}. They found that the disengagement model trained in the single-user condition might not be appropriate for the group condition, but the group condition model generalizes better to the single-user condition. 
A mixed model combining both conditions is a good compromise, but it does not achieve the performance levels of the models trained for a specific type of interaction. Their best models give an accuracy of 63\% and AUC of 0.61 for the single-user condition and an accuracy of 73\% and AUC of 0.62 for the group condition. 
This finding has encouraged us to work with mixed conditions. 
\added[id=aby]{Liu \emph{et al.} applied the Echo State Networks (ESNs) architecture, a variant of Recurrent Neural Networks, to a real-world dataset 
and showed that these networks are able to predict engagement breakdown behavior using 30 seconds of facial expression features \cite{liu2018predicting}.

}
Our positioning in relation to these previous studies is as follows. First, \added[id=aby]{our collected dataset targets the diverse social signals that are involved in user engagement, considering a wide range of heterogeneous sensors: microphone array, cameras, sonars, lasers, along with user tracked variables (\textit{i.e.} face features, head angles, eye gaze and position toward the robot). To the best of our knowledge, none of the existing datasets provide such a thorough coverage of signals amenable to exploitation for user engagement analysis. This is also the first significant dataset offering a large amount of data collected by the robot ``Pepper''. Pepper offers a large combination of features compared to the other robots used in the literature (NAO, iCub, MyKeepon, and so on).} Second, the use of such a large and real-world dataset allows us to investigate deep learning approaches such as recurrent neural networks for the multimodal detection of  user engagement decrease. This ``into-the wild'' dataset is here used to model the temporal user behavior in order to make decision in real-time about engagement decrease. It follows the work of : \textit{i)} \cite{liu2018predicting} that uses such neural networks with facial expression alone on a reduced set of our dataset that has already been made public; \textit{ii)}\cite{Ben-Youssef2019ieee} that shows the superiority of multimodality for a related but close task which is the prediction of engagement breakdown using task-designed logistic regression.
This could lead to the development of lifelike humanoid SAR that could better understand the behavior of the humans they are interacting with, and therefore respond more appropriately in order to increase their engagement. 


\section{Spontaneous Human-Humanoid Interactions}
\label{sec:SpontaneousHRI}

\subsection{Experimental Design} 

\begin{figure}[tb!]
\centering
\includegraphics[width=0.95\linewidth]{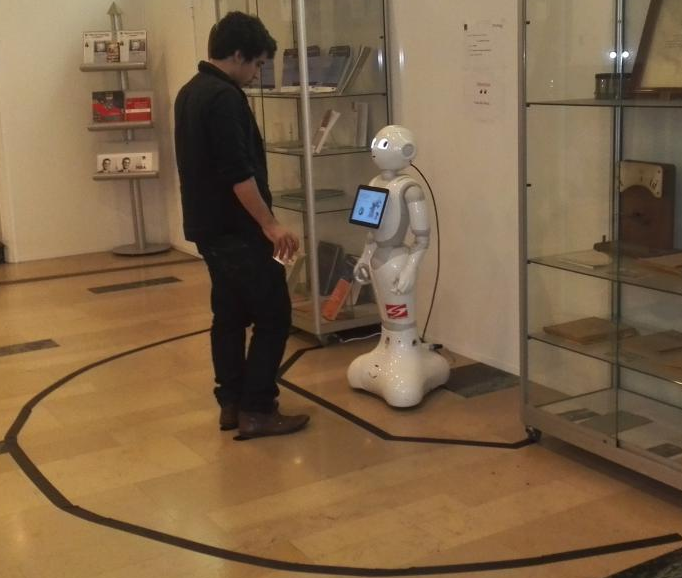}
\caption{Participant in the first engagement zone (less than 1.5 meters from the robot) interacting with Pepper.}
\label{fig:monouser}
\end{figure}

\begin{figure}[tb!]
\resizebox {\columnwidth} {!} {
\begin{tikzpicture}[every 
	node/.style={minimum width=0.4cm},
    leaf/.style={circle,draw,fill=green!20,minimum size=1mm},
    pool/.style={draw,fill=blue!20,minimum width=4cm,minimum height=1cm},
     Arr/.style={
           ->,
           shorten <=2pt,
           shorten >=2pt,},
	]
\node (Pepper) {\includegraphics[width=0.08\columnwidth]{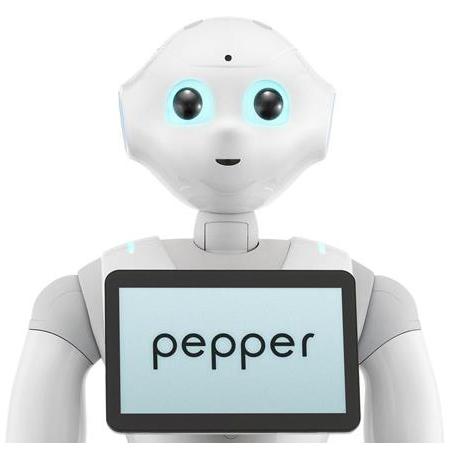}};
\human[0.35, yshift=-23mm]{A}{Pepper}
\human[0.2, gray, xshift=-7.5cm, yshift=-22mm]{B}{Pepper.west}
\human[0.25, gray, xshift=-3.8cm, yshift=-55mm]{C}{Pepper.west}
\human[0.25, gray, xshift=4.9cm, yshift=-45mm]{D}{Pepper.east}
\node[below=of A, yshift=1.05cm] {\textit{{\tiny Interaction zone}}};
\draw[dashed] (-1.6,0) arc (180:360:1.6cm);
\node (server) at (-2.2,0.8) [draw, fill=gray, align=center, text width=0.8cm]{\tiny Storage\\[-1mm] Server};
\draw[Arr] (Pepper)   to [bend left=-30] (server);
\draw (-1.5,0.37) -- node[above] {} ++(3,0.0);
\node[right=of server, xshift=-0.6cm, yshift=1.9mm] {\textit{{\tiny Transfer over Ethernet}}};
\end{tikzpicture}
}
\caption{Technical setup.}
\label{fig:setting}
\end{figure}
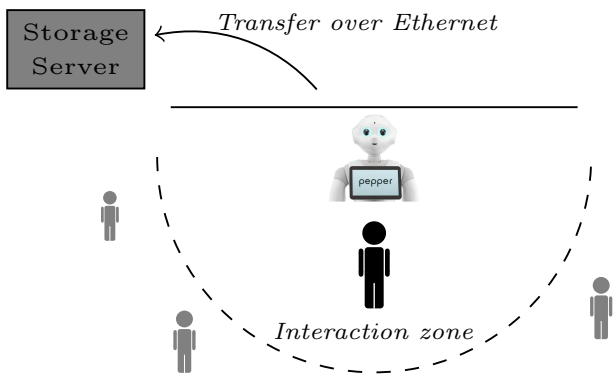

The experiments were conducted in a public space at Telecom ParisTech over 17 months. 
The recordings consisted of interactions between humans and the robot Pepper (see Fig.~\ref{fig:monouser}). The collected data constitute the UE-HRI dataset\footnote{\url{https://www.tsi.telecom-paristech.fr/aao/en/2017/05/18/ue-hri-dataset/}} described in \cite{Ben-Youssef2017}. 
It includes all data streams available on Pepper, packaged in the open-source Robot Operating System (ROS) framework\footnote{\url{http://wiki.ros.org/naoqi\_driver}}. 
Each stream is translated into a message (called ROS topic) and packaged together into a ROSbag file.
In order to keep all the streams synchronized, they were indexed using the robot timestamps. 
The recorded data is split into ROSbag files of 100 Mb in order to quickly move them from the robot to a storage server over Ethernet. ROSbag files are then merged together into one ROSbag file in order to get one file per interaction. Fig.~\ref{fig:setting} shows the experimental setup of the interaction. 

Pepper automatically starts the interaction when it detects movement, and focuses on the participant in front of it, who is in the interaction zone (\textit{i.e.} a distance of less than $1.5$-meters from the robot, indicated by means of black tape stuck on the floor in Fig.~\ref{fig:monouser}). 

First, the robot 
asks the user to sign the \textit{agreement} form displayed on its embedded tablet, authorizing researchers to use her collected data for further analysis. 
After validation of the agreement, the robot enters the \textit{welcome} phase by introducing itself through very lively animations and providing the user with the following instructions: ``speak loud and be alone in the $1^{st}$ engagement zone''. It then enters the \textit{dialog} phase. This includes a set of open-ended questions where the robot asks the participant to introduce herself and to talk about her favorite restaurants and films.
The next phase is the \textit{cucumber} phase, when the robot presents its vision technology to the user in a humoristic way by showing that, from its viewpoint, the difference between a cucumber and a human is the face. Finally, the robot enters the \textit{survey} phase, during which the user is asked to assess her satisfaction with the interaction with Pepper, by answering 15 questions on a 5-level Likert scale (from disagree ``1'' to agree ``5'') \cite{Glas2015,Celiktutan2017} (see ~\nameref{sec:AppendixA}). 

\subsection{Participants}

The recordings involved 278 users ($182$ males, $96$ females)
, whose average age was 25 ($\pm 9.5$) years. This was estimated using an \emph{ad hoc} software module embedded in Pepper \cite{OMRON}.
The users were students, teachers, researchers, visitors and other staff of Telecom ParisTech.
A poster on the wall warned users that they were being recorded during the interaction with the robot. The contact information of the main researcher was also made available on the poster. This was done to allow the users to contact the researcher, should they have concerns about the exploitation of their data, and to be able to ask to have it deleted if they so wished. 
No instructions were given to the user except those provided by the robot in the welcome phase. Users were free to participate in the interaction and free to leave when they wished. 
The interaction was unsupervised, so the number of users simultaneously involved in it was not controlled. Even though the robot warned that only one user was to be in the first engagement zone at a time, the collected data included $209$ interactions featuring a single user, and $69$ multiparty interactions ($32$ started as multiparty and ended as single-user). Note that only 46 users stayed until the end of the scenario and the remaining 72, 84, 70 and 6 users left the interaction at the welcome, dialogue, cucumber and survey phase, respectively. 


\subsection{Social Signals on the Robot Pepper}

Pepper can record a large variety of data streams ranging from raw signals (audio, video, sonar and laser) to face tracking and estimation of gaze direction, head motion and facial expression. 
In this work, features were extracted by using the available trackers of NAOqi-SDK as they are integrated in the robot.

\textbf{Distance:}
The distance between the user and the robot was computed using measured raw signals (i.e. sonar) and tracked variables as described below. More specifically, the front sonar\footnote{\url{http://doc.aldebaran.com/2-7/family/pepper_technical/sonar_pep.html}} (\textit{i.e.} ultrasonic sensor) was used.
The NAOqi People Perception\footnote{\url{http://doc.aldebaran.com/2-7/naoqi/peopleperception/}}  module was also used to extract the distance of the participant's face from the robot camera as well as her 3D head position in relation to the robot’s torso reference.
The space in front of the robot was divided into three configurable zones using the ALEngagementZones module. 
The default configuration was used here. The first engagement zone is the area about 1.5m away from the robot. In this work, this was used as the interaction zone. The second zone is the area between 1.5m and 2.5m away. The third zone 
is the area more than 2.5m away from the robot.
The participant's position was classified to be in one of these three spaces (or 0 if unknown) 
using the 3D coordinates of the user's head in the robot frame\footnote{\url{http://doc.aldebaran.com/2-7/glossary.html#term-frame-robot}}. 

\begin{table}[tb!]
  \caption{Extracted stream feature}
    \label{tab:feature}
\resizebox {\columnwidth} {!} {
\begin{tabular}{l l l}
   \hline
\textbf{Stream} & \textbf{Feature}  & \textbf{Description} \\
   \hline
Distance & Front sonar & 1 feature (in meter)\\
 & Face distance & 1 feature (in meter)\\
 & 3D head position & 3 features [x, y, z] (in torso frame)\\
 & Engagement zone & 1 feature $\in \{0, 1, 2, 3\}$\\
   \hline
Gaze & Gaze direction & 2 features [yaw, pitch] (in radians) \\
 & Is looking at robot & 1 feature $\in \{0, 1\}$\\
   \hline
Head & Head angles & 3 features [yaw, pitch, roll]\\ && (in radian)\\
   \hline
Face 
 & \replaced[id=aby]{Action Unit}{ Facial expression} & \replaced[id=aby]{17}{ 5} features $\in [0,1]$\\
   \hline
Speech & Voicing probability & 1 feature \\
 & F0 & 1 feature \\
 & Loudness & 1 feature \\
 & Log(energy) & 1 feature \\
 & 12 MFCCs & 12 Features \\
 & Is Robot Speaking & 1 feature $\in \{0, 1\}$\\
 & Robot speech duration & 1 feature (in second)\\
 & User speech duration & 1 feature (in second)\\
   \hline
 \end{tabular}
 }
\end{table}



\textbf{Gaze:}
Pepper's ALGazeAnalysis module gives information about the user face orientation in order to detect whether the user is looking at the robot or not. 
OpenFace 2.0 \cite{Baltrusaitis2018} was used to compute gaze direction in relation to the plane of the face \cite{Wood2015}. 

\textbf{
Head and Face:}
OpenFace 2.0 \cite{Baltrusaitis2018} was also used to compute the head pose of the user along the three axes (yaw, pitch, roll). Moreover, it was also used to recognize the occurrence and intensity of each facial Action Unit (AU) \cite{Baltrusaitis2015}.


\textbf{Speech:}
The audio signal was recorded at a sampling frequency of 48KHz 
using 4 microphones that are available inside the head of the robot. 
The audio signal contains the speech of both the participant and the robot 
as well as noise in the environment. 
In order to simplify the analysis of the audio, we selected the first channel (\emph{i.e.} first microphone) to extract speech features\footnote{Beamforming would be a better alternative that will be considered in future work}. Speech features included: the fundamental frequency (F0) (extracted via an autocorrelation and cepstrum based method), log-energy, loudness contours, voicing probability and the first 12 MFCCs excluding the $0^{th}$ MFCC. All these features were computed from the audio signal over 50-ms windows at a frame rate of 100 Hz with openSMILE \cite{Eyben2010}. 
Features indicating if the robot is speaking or not, as well as the robot's and the user's speech duration, were computed from the dialog (Text-To-Speech and Automatic Speech Recognition) ROSbag topics. 

\subsection{Annotation of Engagement}

We developed a script that extracts synchronized front and bottom images\footnote{\url{http://doc.aldebaran.com/2-7/family/pepper_technical/video_2D_pep.html}} and audio from the corresponding ROSbag topics and merges them into a video using ffmpeg\footnote{\url{https://www.ffmpeg.org/}}.
Two annotators with different scientific backgrounds annotated the dataset: a researcher who knew the purpose of the work and an uninformed one who did not.
The ELAN annotation tool \cite{Wittenburg2006} was used to annotate the videos. 
On all recordings, the annotator indicates 
the start and the end of the interaction
as well as the number of participants (\textit{i.e.} mono-user or multi-users).
In order to characterize engagement, 
annotators were asked to annotate the interaction video segment by segment based on verbal and non-verbal behaviors expressed by the user that exhibits an
engagement decrease, with the following label ``\textit{Sign of Engagement Decrease (SED)}''. 

A sign of engagement decrease (SED) reflects any cue exhibited by the user showing any form of disinterest in the robot. It could occur any time during the interaction. This cue may correspond to verbal or nonverbal behaviors of the participant. 
SED could represent an early sign of future engagement breakdown, that is, a sign that leaving the interaction will occur in the near future and before the end of the scenario. 

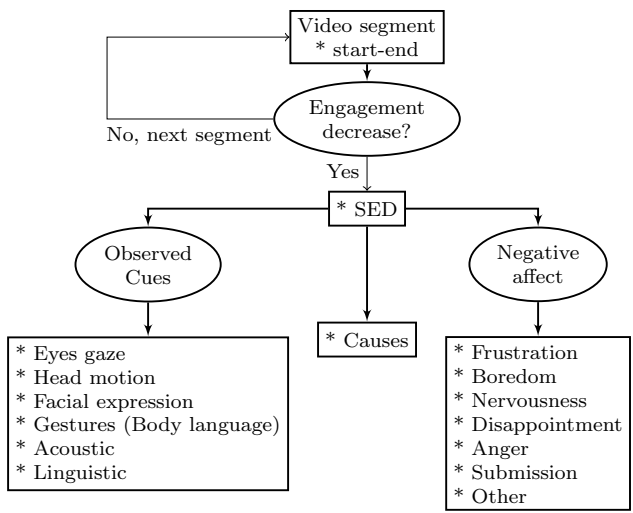
\begin{figure}
\resizebox {\columnwidth} {!} {
    \begin{tikzpicture}[box/.style={draw, ellipse, thick, align=center, minimum height=0.5cm, minimum width=1cm},
  line/.style={draw, thick, -latex'}]
  
        \node [box, rectangle]  (video)      {Video segment\\* start-end};
        \node [box, below=0.25cm of video]              (segment)      {Engagement\\decrease?};
        \node [box, rectangle, below=0.5cm of segment]              (SED)      {* SED}; 
        \node [box, rectangle, below=1.4cm of SED]    (causes)    {* Causes};
        \node [box, below=-0.cm of SED, xshift=-3.2cm]    (Cues)    {Observed\\Cues};
        \node [box, below=-0.cm of SED, xshift=2.5cm]     (affect)    {Negative\\affect};
        \node [box, rectangle, below=0.5cm of affect, align=left]  (negaffect)    {* Frustration\\
* Boredom\\
* Nervousness\\
* Disappointment\\
* Anger\\
* Submission\\
* Other};
        %
        \node [box, rectangle, below=0.5cm of Cues, align=left]   (cueslist)    {* Eyes gaze\\
* Head motion\\
* Facial expression\\
* Gestures (Body language)\\
* Acoustic\\
* Linguistic};
        \path [line] (video) --        (segment);
\draw[->] (segment) -- node[below right,pos=1.05]{No, next segment} ++(-3.8,0) |- (video);     
\draw[->] (segment) -- node[below left,pos=0.0]{Yes} (SED);
        \path [line] (SED) -|         (Cues);
        \path [line] (SED) --         (causes);
        \path [line] (SED) -|       (affect);
        \path [line] (Cues) --       (cueslist);
        \path [line] (affect) --       (negaffect);
    \end{tikzpicture}
}
\caption{Flow-chart of the different annotation levels. The '*' shows what the annotator has to select.}
\label{fig:annotation}
\end{figure}

Fig.~\ref{fig:annotation} shows a flow-chart that summarizes the annotation process described above. 
\added[id=aby]{A video tutorial was created to explain the annotation process and how to annotate the interaction using ELAN.}
The annotator defines the start and the end segment as well as the corresponding label, observed cues and negative affect of that segment.
For each defined segment, the annotator assigns the corresponding observed cues of that decrease, in order of importance. This part could be sub-segmented. For example, if the participant says:``I'm bored'', with a corresponding facial expression, the annotator indicates in the ``Cues 1'' track: ``speech linguistic'' and in ``Cues 2'' track: ``face''. The annotator decides which one is more visible in the segment to appear in ``Cues 1''. If these two cues are successive in time, both should appear in ``Cues 1'' with a sub-segmentation of the start and end of each one.
The annotator also assigns the corresponding negative affect of that segment (if relevant) of that decrease. Negative affects (frustration, boredom, nervousness, disappointment, anger, submission) are based on verbal and nonverbal behavior while interacting with Pepper. 
Annotators are free to add more information concerning this segment. We recommend that they add information about the causes in the ``Causes'' track. 

\begin{figure}[h]
  \centering
  \subfloat[Annotation of the whole interaction 
  $\kappa=0.73$]{\label{fig:allinteraction}\includegraphics[width=\columnwidth,trim=0 0 0 0, clip]{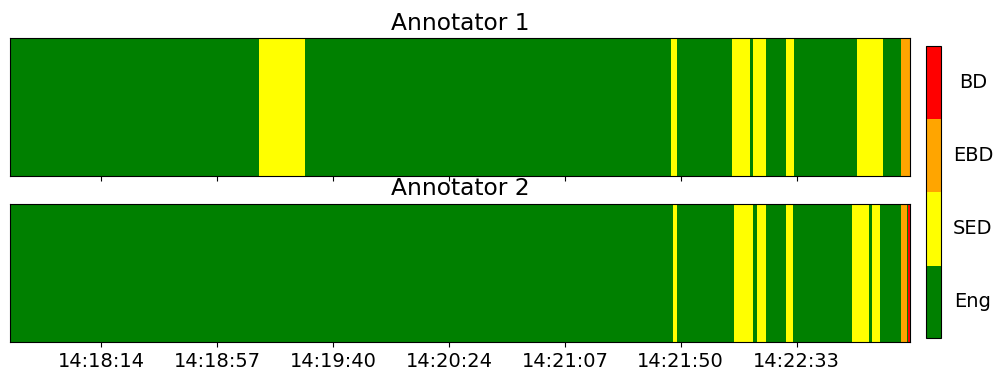}}\\
  \subfloat[Zoom on the last 2 min of the interaction ($\kappa=0.73$)]{\label{fig:interaction2min}\includegraphics[width=\columnwidth,trim=0 0 0 0, clip]{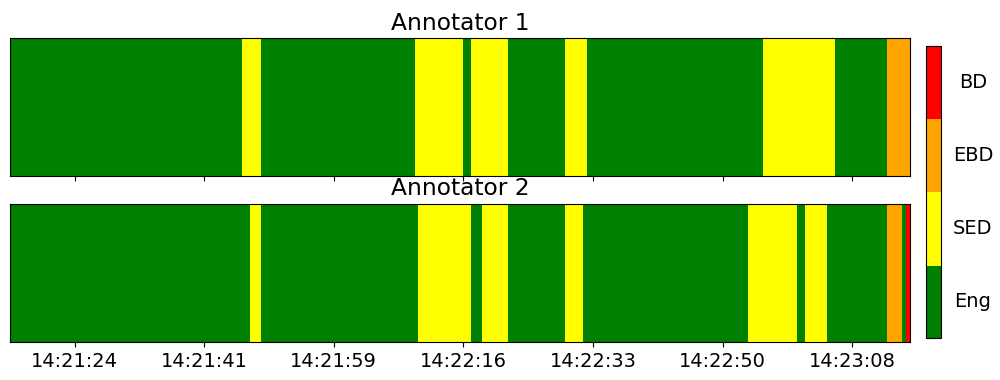}}\\
  \subfloat[Zoom on the last 2 min of the interaction with ignoring the ``Engaged'' segment less than 1 second located between 2 SED segments. $\kappa=0.74$]{\label{fig:interaction2min_ignored1sec}\includegraphics[width=\columnwidth,trim=0 0 0 0, clip]{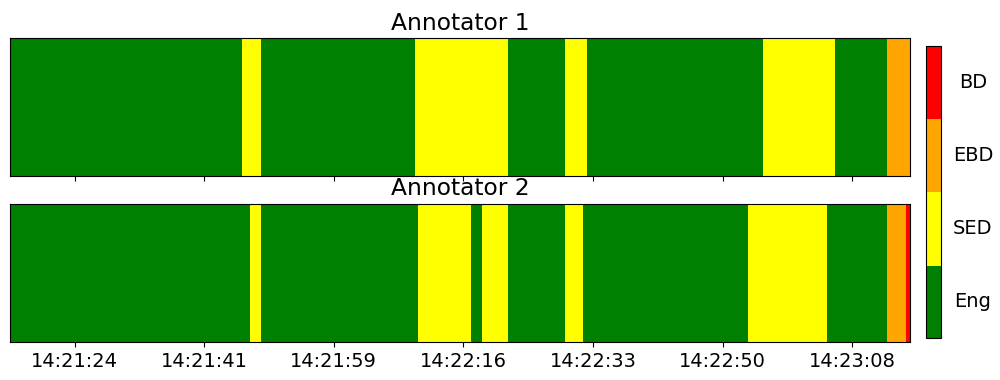}}\\
  \caption{Example of annotation. BD: Engagement BreakDown \textit{i.e.} leaving before the end of the interaction scenario. EBD: early sign of engagement breakdown (EBD) (\textit{i.e.} the last SED of the interaction that BD will occur just after).}
  \label{fig:example_annotation}
\end{figure}

The overall \textit{Cohen kappa} agreement score on annotated recordings for SED annotation is $\kappa=0.73$ (substantial agreement) (see Fig.~\ref{fig:example_annotation} in \nameref{sec:AppendixB}). 
If we automatically correct the annotations by merging together the ``engaged'' segment located between 2 SED segments and inferior to 1 second in duration, to get 1 large SED segment instead of 2 separated by 1 or 2 frames of ``engaged'', the Kappa increases slighty to $\kappa=0.74$ (see Fig.~\ref{fig:example_annotation}).

\section{User Engagement Decrease Analysis}
\label{sec:analysis}

\begin{figure}[tb!]
  \centering
  \subfloat[Cues]{\rotatebox[origin=l]{90}{\hspace{4em}\#Occurrences}
  \label{fig:cues}\includegraphics[width=\columnwidth,trim=0 0 0 16, clip]{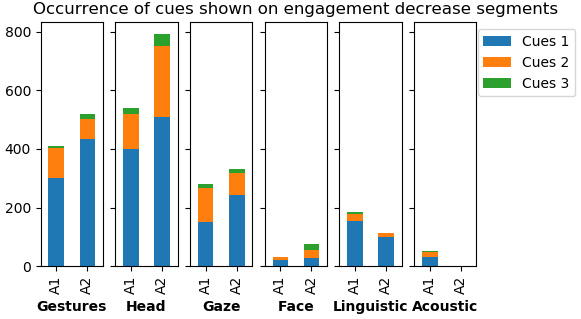}}\\
  \subfloat[Negative affects. The meaning of B: Boredom, F: Frustration, D: Disappointment , N: Nervousness, A: Anger, S: Submission, O: Other]{\rotatebox[origin=l]{90}{\hspace{4em}\#Occurrences}
  \label{fig:affect}\includegraphics[width=\columnwidth,trim=0 0 0 16, clip]{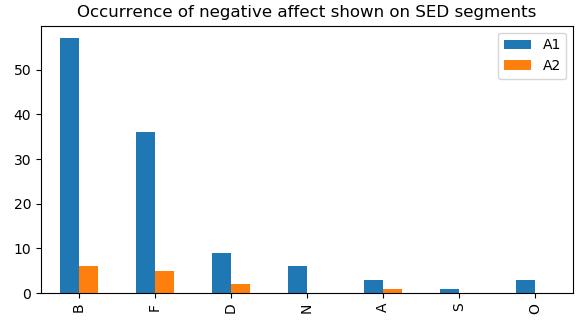}}
  \caption{Cues and affects distribution of signs of engagement decrease (SED) by each annotator. A1: denotes the first annotator 1, A2: denotes the second annotator.} 
  \label{fig:annotated_sued}
\end{figure}

%

\begin{figure*}
  \centering
  \subfloat[Distance] 
  {\label{fig:distance}\includegraphics[width=0.5\textwidth]{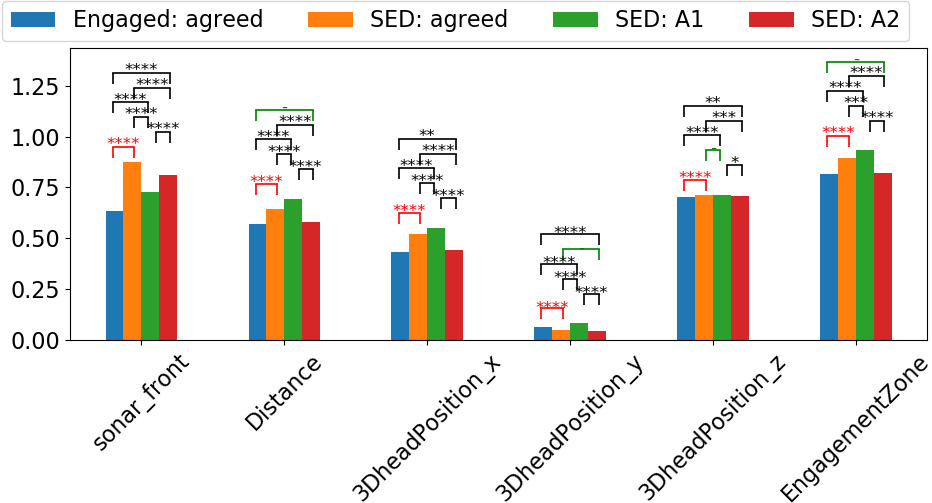}\label{fig:Distance}}
  \subfloat[Gaze] 
  {\label{fig:gaze}\includegraphics[width=0.25\textwidth]{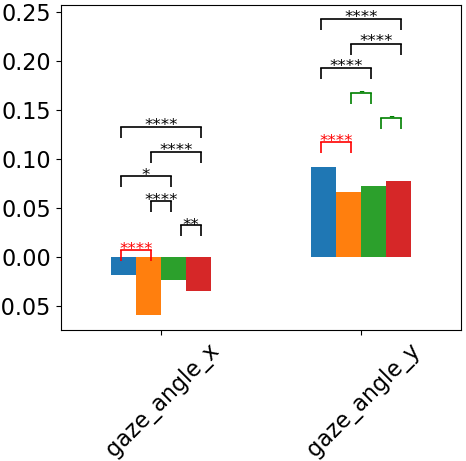}}
  \subfloat[Head motion variation] 
  {\label{fig:head}\includegraphics[width=0.25\textwidth]{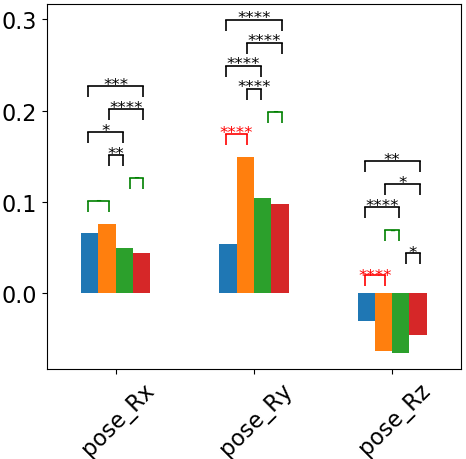}} \\
  \subfloat[Facial expression] {\label{fig:au}\includegraphics[width=0.5\textwidth]{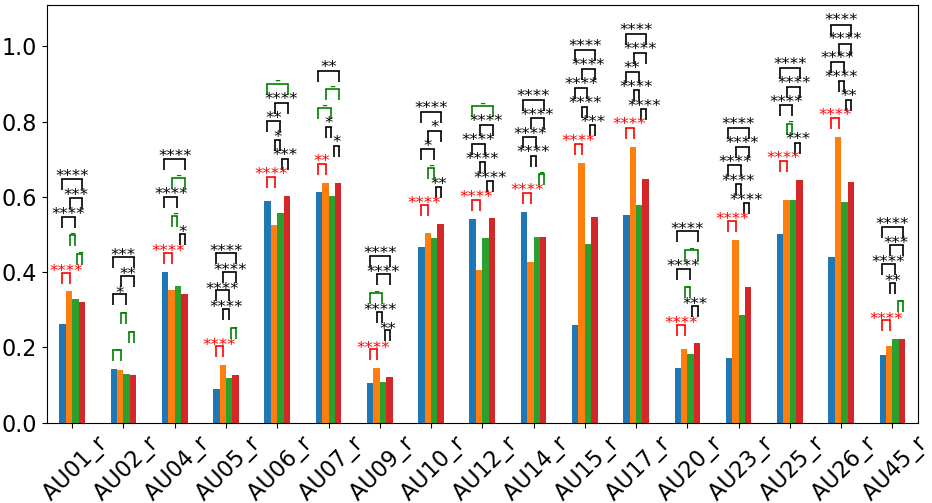}}
\subfloat[Looking and listening] {\label{fig:look}\includegraphics[width=0.25\textwidth]{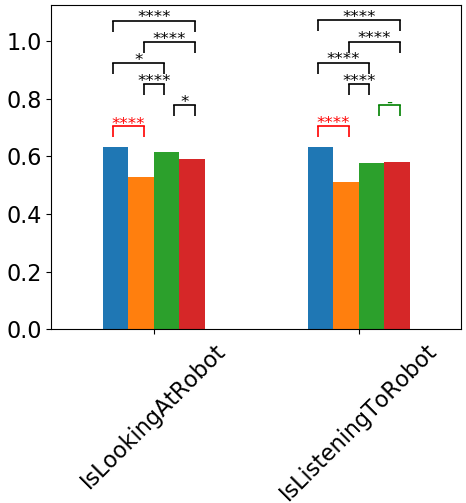}}
\subfloat[Speech duration] {\label{fig:voice}\includegraphics[width=0.25\textwidth]{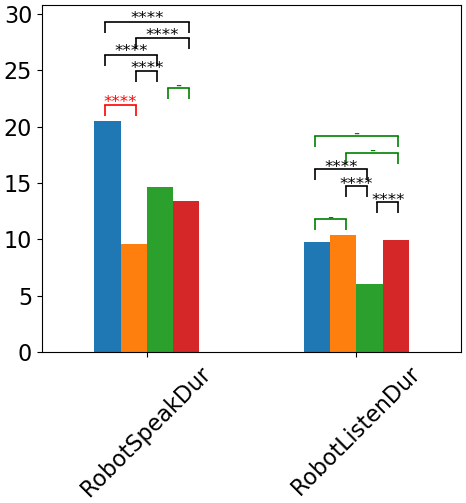}}
  \caption{Selected features of users' behavior when the two annotators (A1 and A2) agreed on their engagement as well as when they disagreed. Paired T-tests were calculated: red was used when the annotators agreed, **** means $p < 0.0001$, *** means $p < 0.001$, ** means $p < 0.01$, * means $p < 0.05$ and green ``-'' means $p >= 0.05$.
   }
  \label{fig:behavior}
\end{figure*}



According to both annotators, the average duration of the interaction is 7 ($\pm 5$) minutes. During interactions, users displayed SED in around 6 segments lasting in average 6 ($\pm 9$) seconds. \added[id=aby]{Note that for participants who left the interaction by the second phase, the intervals between SED were shorter compared to those who stayed till the end of the interaction.}
Note \added[id=aby]{also} that the last segment where SED were shown is generally longer. 
In average, its duration is around 9 ($\pm 15$) seconds. 
In 90\% of the interaction duration, the users are engaged. For the reaming 10\%, the users exhibit SED.

Fig.~\ref{fig:annotated_sued} displays the \replaced[id=aby]{number of occurrences}{ statistics} of the behavior exhibited by the users when their engagement decreases, as perceived by the annotators. 
Fig.~\ref{fig:cues} confirms that the non-verbal behaviors play a special role to point the engagement level. 
Head motion, gesture (\emph{i.e.} posture, hand waving, and so on) and eye gaze are the most recurrent features to identify a decrease of engagement in our dataset. 
Fig.~\ref{fig:affect} shows that annotators disagreed 
on selecting the appropriate affects related to the SED segments. This showed that the annotation of affects was more subjective here than the annotations of the SED category and their cues. 

\begin{table}[tb!]
\caption{Engagement decrease causes}
\label{tab:Causes}       
\begin{tabular}{ll}
\hline\noalign{\smallskip}
Causes of SED & Rate (\%)  \\
\noalign{\smallskip}\hline\noalign{\smallskip}
User interrupted by another person & 39\% \\
Robot error (long pauses, misunderstood) & 17\% \\
User uses his phone & 10\% \\
Robot focus on another person & 5\% \\
User time constraint & 2\% \\
User missed robot's request & 2\% \\
\noalign{\smallskip}\hline
\end{tabular}
\end{table}

Due to the wide variety of possible factors that can cause engagement decrease in spontaneous interactions, it is difficult to determine the exact cause for each SED segment.
However, we asked the annotators to try to mention any information related to the cause of that decrease. 
Table~\ref{tab:Causes} presents the main causes of the engagement decrease detected by the two annotators, with their percentage of occurrence. 
We individuated two principal sources that lead to the decrease of engagement: the first is due to a social norm violation 
(\emph{e.g.} another person interrupts the interaction while the robot is talking; user time constraint; user is using her phone); the second cause is due to robot's technical issues (\emph{e.g.} robot makes long pauses or misunderstands the user). 

We compared users' behaviors when they were engaged with the robot vs. when they showed signs of engagement decrease based on the annotations. 
Figure~\ref{fig:behavior} presents the results of the comparison for the different configurations: when both annotators perceived the user as being engaged (denoted by ``Engagement agreed''), when both annotators agreed about the user engagement decrease (denoted by ``SED agreed'') and when both annotators disagreed about the engagement state (denoted by ``SED: A\textit{x}'' when a decrease of engagement is perceived by one annotator \textit{x} and not by the other one).
Figure~\ref{fig:distance} shows the average distance between the user and the robot. The users were closer to the robot when they were fully engaged than when their engagement decreased. 
Regarding gaze, when users were engaged they looked more at the robot than when their engagement decreased (Figure~\ref{fig:look} ``1'' when the user looks at the robot, ``0'' otherwise). This could be confirmed with vertical gaze direction around pitch axis (\emph{i.e.} angle x) in Figure~\ref{fig:gaze}. 
Head motion (\emph{i.e.} shaking, tilting and nodding) were displayed in Figure~\ref{fig:head}.
Users move their head more when their engagement decreases. 
Concerning action units (AU) \cite{Rawassizadeh2016} (see Figure~\ref{fig:au}), we found that users have the appearance of being happier (where happiness involves AU06 and AU12) 
when they are engaged, compared to when their engagement decreases. 
Similarly, for sadness, which is the combination of AU01, AU04, AU15, anger (the combination of AU04, AU05, AU07, AU23) and disgust (the combination of AU09, AU15, AU16), it appears that users express these negative emotions when their engagement decreases, compared to when they are engaged. 
Figure~\ref{fig:voice} shows that the users are more engaged when the robot is speaking. This could be confirmed with Figure~\ref{fig:look} (\emph{i.e.} ``1'' when the robot is speaking, ``0'' when the robot is listening).

In the next section, for training and testing of engagement decrease detection, we consider only the segments 
where both annotators agreed on the engagement category. 

\section{Detection of User Engagement Decrease}
\label{sec:modelling}

\subsection{User Engagement Modeling}
\label{ssec:problemspecification}

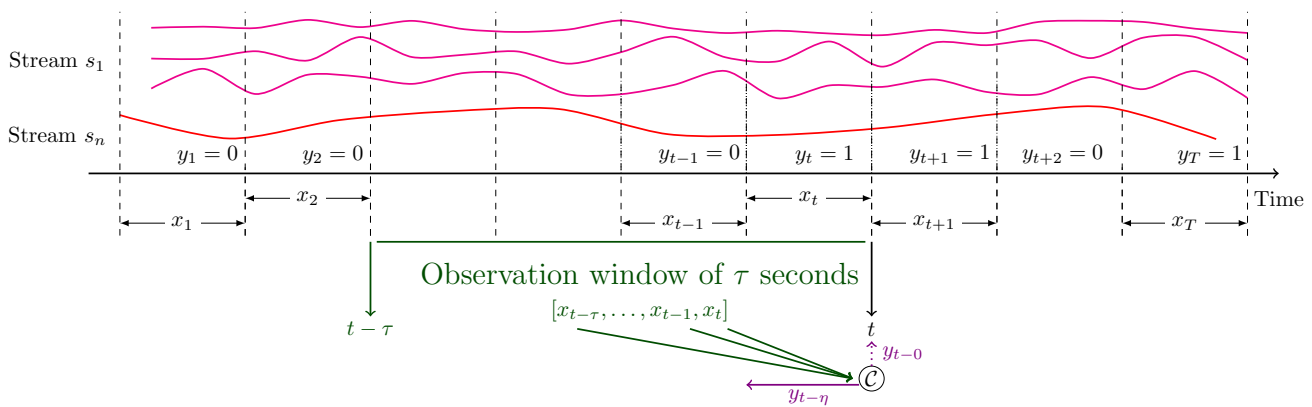
\begin{figure*}
\resizebox {\textwidth} {!} {
\begin{tikzpicture}[font=\fontsize{10}{10}\selectfont,
  doublearrow/.style={
    <->, 
    >=latex,
   every node/.style={fill=white,font=\fontsize{10}{10}}}]
\pgfmathsetseed{42}
\pgfmathsetmacro\start{1}
\pgfmathsetmacro\fin{20}
\pgfmathsetmacro\step{2}
\draw [thick,->] (\step-0.5,0) -- (\fin+0.5,0);
\node[align=center] at (\fin+0.5,-0.4) {Time};
\draw (1,1.8) node {Stream $s_1$};
\draw [thick,magenta] plot[smooth,domain=\step+0.5:\fin,samples=2*(\fin/\step+1)] (\x,rnd*0.5+1.2);
\draw [thick,magenta] plot[smooth,domain=\step+0.5:\fin,samples=2*(\fin/\step+1)] (\x,rnd*0.5+1.7);
\draw [thick,magenta] plot[smooth,domain=\step+0.5:\fin,samples=2*(\fin/\step+1)] (\x,rnd*0.3+2.2);
\draw (1,0.6) node {Stream $s_n$};
\draw [thick,red] plot[smooth,domain=\step:\fin-0.5,samples=\fin/\step+1] (\x,rnd*0.6+0.5);

\pgfmathsetmacro\nbsamples{\fin/\step-1}
\foreach \x in {\start, ..., \nbsamples}{
    \draw[dashed]  (\x*\step,-1) -- ++ (0,3.6);
}
\foreach \x in {\start, ..., 2}{
    \draw[dashed] (\x*\step+\step,-1) -- ++ (0,1.1) node [xshift=-6mm,yshift=2mm]{$y_{\x}=0$};
}
\pgfmathsetmacro\lastsamples{\nbsamples-2}
\pgfmathsetmacro\x{\lastsamples-1}
    \draw[dashed] (\x*\step,-1) -- ++ (0,1.1) node [xshift=-7.5mm,yshift=2mm]{$y_{t-1}=0$};
\pgfmathsetmacro\x{\lastsamples}
    \draw[dashed] (\x*\step,-1) -- ++ (0,1.1) node [xshift=-7.5mm,yshift=2mm]{$y_{t}=1$};
\pgfmathsetmacro\x{\lastsamples+1}
    \draw[dashed] (\x*\step,-1) -- ++ (0,1.1) node [xshift=-7.5mm,yshift=2mm]{$y_{t+1}=1$};
\pgfmathsetmacro\x{\lastsamples+2}
    \draw[dashed] (\x*\step,-1) -- ++ (0,1.1) node [xshift=-9.5mm,yshift=2mm]{$y_{t+2}=0$};
\draw[dashed] (\fin,-1) -- ++ (0,3.6) node [xshift=-6mm,yshift=1mm]{}; 
\draw[dashed] (\fin,-1) -- ++ (0,1.1) node [xshift=-6mm,yshift=2mm]{$y_{T}=1$};

\pgfmathsetmacro\FrameStart{\start+\nbsamples+\step}
\pgfmathsetmacro\FrameW{\step}
\pgfmathsetmacro\FrameOverlap{0.0}
\pgfmathsetmacro\FrameStep{\FrameW-\FrameOverlap}

\foreach[
  evaluate=\i as \y using {ifthenelse(mod(\i,2)==0,-0.4,-0.8)},
  evaluate=\i as \X using \i*\FrameStep,
  count=\j
] \i in {\start, ..., 2} {
  \draw [doublearrow] (\X,\y) -- node{$x_{\i}$} (\X+\FrameW,\y);
}
\draw[doublearrow] (\fin-\FrameW,-0.8) -- node{$x_{T}$} (\fin,-0.8);

\foreach[
  evaluate=\i as \y using {ifthenelse(mod(\i,2)==0,-0.4,-0.8)},
  evaluate=\i as \X using \i*\FrameStep+\FrameStart,
  count=\j
] \i in {-1} {
  \draw [densely dotted] (\X,0) -- (\X,1.5);
  \draw [densely dotted] (\X+\FrameW,0) -- (\X+\FrameW,1.5);
  \draw [doublearrow] (\X,\y) -- node{$x_{t-1}$} (\X+\FrameW,\y);
}
\foreach[
  evaluate=\i as \y using {ifthenelse(mod(\i,2)==0,-0.4,-0.8)},
  evaluate=\i as \X using \i*\FrameStep+\FrameStart,
  count=\j
] \i in {0} {
  \draw [densely dotted] (\X,0) -- (\X,1.5);
  \draw [densely dotted] (\X+\FrameW,0) -- (\X+\FrameW,1.5);
  \draw [doublearrow] (\X,\y) -- node{$x_{t}$} (\X+\FrameW,\y);
}
\foreach[
  evaluate=\i as \y using {ifthenelse(mod(\i,2)==0,-0.4,-0.8)},
  evaluate=\i as \X using \i*\FrameStep+\FrameStart,
  count=\j
] \i in {1} {
  \draw [densely dotted] (\X,0) -- (\X,1.5);
  \draw [densely dotted] (\X+\FrameW,0) -- (\X+\FrameW,1.5);
  \draw [doublearrow] (\X,\y) -- node{$x_{t+1}$} (\X+\FrameW,\y);
}

\pgfmathsetmacro\place{-1.7}
\pgfmathsetmacro\observation{3}
\node[font=\fontsize{14}{14}\selectfont,align=center, black!70!green] at (-\observation+\FrameStart+1.3,-3.3-\place) {Observation window of $\tau$ seconds};
\draw [thick, black!70!green] (-\observation*\FrameStep+\FrameStart+0.1,-2.8-\place) -- (\FrameStep+\FrameStart-0.1,-2.8-\place);
\node[align=center, black!70!green] at (-\observation+\FrameStart+1.3,-3.9-\place) {$[x_{t-\tau}, \dots, x_{t-1}, x_t]$};

\draw (\FrameStep+\FrameStart,-5-\place) circle (.2cm);
\node[align=center] at (\FrameStep+\FrameStart,-5-\place) {$\mathcal{C}$};

\draw [thick,->, black!70!green] (-\observation*\FrameStep+\FrameStart,-1.1) -- (-1*\observation*\FrameStep+\FrameStart,-4-\place);
\node[align=center, black!70!green] at (-1*\observation*\FrameStep+\FrameStart,-4.2-\place) {$t-\tau$};
\draw [thick,->] (\FrameStep+\FrameStart,-1.1) -- (\FrameStep+\FrameStart,-4-\place);
\node[align=center] at (\FrameStep+\FrameStart,-4.2-\place) {$t$};

\pgfmathsetmacro\pred{\nbsamples-\lastsamples+\step+1}
\draw [thick,->, black!70!green] (\FrameStart+-1*\FrameStep-0.7,-4.2-\place) -- (\FrameStep+\FrameStart-0.3,-5-\place);
\draw [thick,->, black!70!green] (\FrameStart+-0.1*\FrameStep-0.7,-4.2-\place) -- (\FrameStep+\FrameStart-0.3,-5-\place);
\draw [thick,->, black!70!green] (\FrameStart+0.2*\FrameStep-0.7,-4.2-\place) -- (\FrameStep+\FrameStart-0.3,-5-\place);

\draw [thick,->, dotted,violet] (\FrameStep+\FrameStart,-4.8-\place)--(\FrameStep+\FrameStart,-4.4-\place);
\node[align=center, violet] at (\FrameStep+\FrameStart+0.5,-4.6-\place) {$y_{t-0}$}; 
\draw [thick,->, violet] (\FrameStep+\FrameStart-0.2,-5.1-\place)--(\FrameStep+\FrameStart-\step,-5.1-\place);
\node[align=center, violet] at (\FrameStep+\FrameStart-0.5*\step,-5.3-\place) {$y_{t-\eta}$};

\end{tikzpicture}
}
\caption{Illustration of the detection approach. 
Input: observation window of user behavior is shown in green. 
Output: buffer duration is shown in violet. 
}
\label{fig:detectionapproach}
\end{figure*}

We modeled the task of user engagement decrease detection as a binary classification, where the goal is to predict, in real-time, whether the user is engaged or not with the robot, based on the user's behavior analysis.

Our SED detection approach is illustrated in Figure~\ref{fig:detectionapproach}.
We define \textit{observation window} as a window of $[t-\tau, t]$, that is, a window that ends at time $t$ and takes into account the last $\tau$ seconds of user behavior. We use 
$[x_{t-\tau}, \dots, x_{t-1}, x_t]$ as a feature vector computed over the 
frames of the observation window as input for the classifier. 
As for the output, each observation window is labeled as either engaged or not. 

At running time $t$, we build a model that classifies the observed behavior over $[t-\tau,t]$ as \emph{user engaged} or \emph{user not engaged}. 
Let 
$X = [x_1, x_2, \dots, x_T]$ 
denote the sequence of 
multimodal user-behavior 
feature vectors 
and 
$Y^{\eta} = [y^{\eta}_1,y^{\eta}_2, ..., y^{\eta}_T]$ denote the corresponding sequence of (binary valued) output labels, where $\eta$ is the duration of the buffer for holding more observations and 
\begin{equation}
\label{eq:bd}
y^{\eta}_{t} = \mathcal{C}([x_{t-\tau}, \dots, x_{t-1}, x_t]) \mbox{ \textit{with} } \tau \geq \eta
\end{equation}
where $\mathcal{C}(.)$ 
is the classifier decision function and
$$
\left\{
    \begin{array}{ll}
        y^{\eta}_{t} = 1, & \mbox{\textit{SED perceived at time }}t-\eta \\
        y^{\eta}_{t} = 0, & otherwise\\
    \end{array}
\right.
$$

\subsection{Deep Networks}

In this study, a sequential modeling approach is proposed to detect SED using deep learning techniques \cite{Bengio2009}.

Remembering information for long periods of time is the default behavior of Long-Short Term Memory (LSTM) \cite{Hochreiter1997}. \replaced[id=aby]{LSTM}{They} uses a memory unit that can remember information/context from the beginning of the input sequence (\emph{i.e.} $t-\tau$).
Gated Recurrent Unit (GRU) networks \cite{Cho2014} are similar to the LSTM, but use a simplified structure. Both LSTM and GRU can be used for modeling temporal sequences. 
However, GRU involves less computation units than LSTM, since they do not have an output gate. Therefore LSTM are usually preferred if trained on very large datasets (big data).

\subsection{Experiments}
\label{sec:exp}

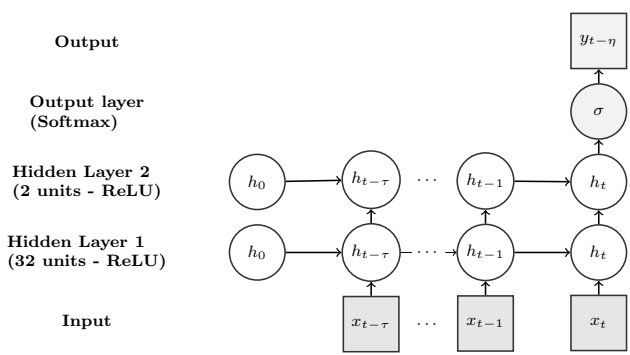
\begin{figure}
\resizebox {\columnwidth} {!} {
\begin{tikzpicture}[main/.style={circle, minimum size = 10mm, thick, draw =black!80, node distance = 10mm},
  connect/.style={thick},
  box/.style={rectangle, minimum size = 10mm, draw=black!100, node distance = 10mm}
]
  \node[box,draw=white!100] (Label) {\textbf{Output}};
  \node[box,draw=white!100, below=0.25cm of Label] (OutputLayer) {\textbf{\makecell[l]{Output layer \\(Softmax)}}};
  \node[box,draw=white!100, below=0.25cm of OutputLayer] (Layer2) {\textbf{\makecell[l]{Hidden Layer 2 \\(2 units - ReLU)}}};
  \node[box,draw=white!100, below=0.25cm of Layer2] (States) {\textbf{\makecell[l]{Hidden Layer 1 \\(32 units - ReLU)}}};
   \node[box,draw=white!100,below=0.25cm of States] (Observed) {\textbf{Input}};
 \node[main] (s0) [right=of States] {$h_{0}$};
  \node[main] (s1) [right=of s0] {$h_{t-\tau}$};
  \node[main] (s3) [right=of s1] {$h_{t-1}$};
  \node[main] (st) [right=of s3] {$h_{t}$};
 \node[main] (s10) [above=0.25cm of s0] {$h_{0}$};
  \node[main] (s11) [above=0.25cm of s1] {$h_{t-\tau}$};
  \node[main] (s13) [above=0.25cm of s3] {$h_{t-1}$};
  \node[main] (s1t) [above=0.25cm of st] {$h_{t}$};
  \node[main, fill=black!5] (pred) [above=0.25cm of s1t] {$\sigma$};
  \node[main,rectangle, fill=black!5] (yt) [above=0.25cm of pred] {$y_{t-\eta}$};
  \node[main,rectangle, fill=black!10] (x1) [below=0.25cm of s1] {$x_{t-\tau}$};
  \node[main,rectangle, fill=black!10] (x3) [below=0.25cm of s3] {$x_{t-1}$};
  \node[main,rectangle, fill=black!10] (xt) [below=0.25cm of st] {$x_{t}$};
  \path (s1) -- node[auto=false]{\ldots} (s3);
  \path ([xshift=.0cm]s1.east) edge ([xshift=-.8cm]s3.west); 
  \path ([xshift=.8cm]s1.east) edge [->] ([xshift=.0cm]s3.west); 

  \path (x1) -- node[auto=false]{\ldots} (x3);
  \path (s1t) edge [connect, ->] (pred);
  \path (pred) edge [connect, ->] (yt);
  \path (s0) edge [connect, ->] (s1)
        (s3) edge [connect, ->] (st); 
  \path (s10) edge [connect, ->] (s11)
        (s13) edge [connect, ->] (s1t); 
  \path (s11) -- node[auto=false]{\ldots} (s13);
  \path (s1) edge [connect, ->] (s11);
  \path (s3) edge [connect, ->] (s13);
  \path (st) edge [connect, ->] (s1t);
  \path (x1) edge [connect, ->] (s1);
  \path (x3) edge [connect, ->] (s3);
  \path (xt) edge [connect, ->] (st);

\end{tikzpicture}
}
\caption{Many-to-One deep architecture with 2 layers.}
\label{fig:deep}
\end{figure}

The data streams of different sampling frequencies were indexed using the robot's timestamps.
To obtain synchronized feature vectors, temporal integration \cite{Joder2009} (a.k.a temporal pooling), is performed over all feature streams using common \textit{integration windows}. 
%
The integrated features are obtained by applying an integration function $f$ over sliding (possibly overlapping) integration windows of length $L$ seconds. 
The functions $f$ used in this study are statistics, namely the mean and variance. 
Also, we fix the integration window length $L$ to 500 ms. No overlapping was used.
%
It was shown that combining multiple features gives the highest performance in disengagement prediction (\emph{c.f.} Section~\ref{ssec:engagementHRI}). Therefore, the synchronized texture-window level feature vectors of Distance, Gaze, Head, Face and Speech Streams shown in Table~\ref{tab:feature} were concatenated together to describe users' behavior and were employed as the input features for the SED detection model. 
\added[id=aby]{Further details are given in our previous work \cite{Ben-Youssef2019ieee}}

Our dataset contains missing values. For example, we have missing values on the face features (\emph{i.e.} head motion, gaze, AU) when some occlusion occur. This happens for instance 
when the robot's head is moving, causing the user's face to go out of the cameras' field of view.
We chose to replace the missing values by means of the corresponding feature from the training data. 
We then normalized the data by subtracting the mean value and dividing by the standard deviation of each feature, using the training data.

The whole dataset using both single user and multiparty interactions was used, since this was reported to be a good compromise in \cite{Leite2015}.  
We used 3-fold cross-validation to train and test a set of SED classifiers. 
The split of train and test sets was done at the interaction-level. Hence, the users of the test set (\textit{i.e.} all observations of the user) were not seen during the training phase, which resulted in a user-independent detection model.

We used scikit-learn's \cite{Scikit-learn} implementation of logistic regression as a baseline and Keras's \cite{chollet2015} implementation for DNN, GRU and LSTM. 
We leave the further optimization of the classifiers' hyper-parameters for future work and focus here on the validation of the usefulness of the recurrent network architectures considered. 

Following preliminary experiments, we used 2 layers with 32 units followed by 2 units, ReLU activation, dropout with probability of 0.1 and the RMSprop algorithm as optimizer to train the deep networks (see Fig.~\ref{fig:deep}).
We used 10\% of the training data as a validation set. We trained each model with 100 epochs, using an early stopping callback to stop the training once the validation accuracy started to decrease. In general, the models converge after a maximum of 35 epochs. 
For logistic regression, we used $\ell_2$ regularization and the inverse of the regularization strength $C$ set to 1.
\added[id=aby]{To deal with the imbalanced data distribution, the weights for each class were computed and used for training the models.}

\subsection{Evaluation criteria}
\label{sec:eval}

Traditionally, the accuracy rate and F1-score have been the most commonly used evaluation criteria. However, they are not well suited to our study because the dataset is unbalanced. 
We have around 90\% of the data labeled as engaged and only 10\% of SED. 
In case of imbalanced data, the accuracy reflects only the underlying class distribution, not the prediction performance of the minority class. 
In order to compute meaningful accuracy and F1-score, the test set should represent the true distribution of both classes. 
Therefore, the test set is resampled to be the average over all the samples of the minority class 
and the $n$-differing samples of the majority class selected from the 
available samples. 
%
%
We also computed the 
area under the receiver operating characteristic curve (AUC) in order to determine which of the models used predicts the classes best.
The AUC corresponds to the probability of correctly identifying the SED class \cite{AndrewP.Bradley1997}. The closer the AUC comes to 1, the \replaced[id=aby]{more accurate}{better} it is. 

\subsection{Results}
\label{ssec:res}

\begin{figure}[tb!]
  \centering
  \includegraphics[width=\columnwidth,trim=0 0 0 0, clip]{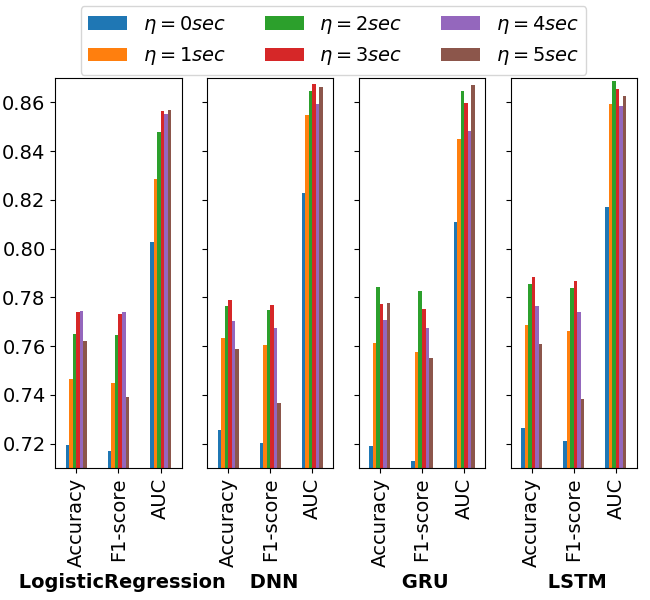}
\caption{Performance of a set of classifiers using 
an observation window of $\tau=5$ sec and a buffer of $\eta \in [0,5]$ sec, respectively.}
\label{fig:classifiers}
\end{figure}

The performance of the different sets of classifiers was compared (see Fig.~\ref{fig:classifiers}). 
We found that deep learning techniques are better 
than conventional machine learning techniques (\emph{i.e.} Logistic Regression).
With the chosen hyper-parameter values the best results were obtained with LSTM for all tested buffer durations $\eta$. 
This is because they better model the temporal dynamics through connections between hidden units in the same layer. 

When we use a buffer delay in the range of $[1,3]$ seconds, the performance of all the classifiers increases. 
This could be explained by the fact that using more information about the user’s behavior plays an important role in inferring the state of their engagement. A buffer longer than 3 seconds does not give a better performance. In addition, a buffer of 3 seconds is already large for real-time detection \cite{Miller1968}. 

\begin{table*}
\caption{LSTM performance for different observation windows $\tau$ (sec) for each buffer $\eta$ (sec) with $\tau \geq \eta$.}
\resizebox {\textwidth} {!} {
    \begin{tabular}{c| c c c c c c | c c c c c c | c c c c c c}
\hline
& \multicolumn{6}{c|}{Accuracy} & \multicolumn{6}{c|}{F1-score} & \multicolumn{6}{c}{AUC} \\
\hline
\diagbox{$\tau$}{$\eta$} & 0 & 1 & 2 & 3 & 4 & 5 & 0 & 1 & 2 & 3 & 4 & 5 & 0 & 1 & 2 & 3 & 4 & 5 \\
\hline
0 & 73.42 & - & - & - & - & - & 0.732 & - & - & - & - & - & 0.820 & - & - & - & - & - \\
1 & \textbf{74.03} & 76.97 & - & - & - & - & \textbf{0.738} & 0.768 & - & - & - & - & \textbf{0.827} & 0.851 & - & - & - & - \\
2 & 73.70 & 76.73 & 77.26 & - & - & - & 0.734 & 0.765 & 0.771 & - & - & - & 0.823 & 0.850 & 0.849 & - & - & - \\
3 & 73.74 & \textbf{77.25} & 78.06 & 78.32 & - & - & 0.734 & \textbf{0.770} & 0.779 & 0.782 & - & - & 0.826 & \textbf{0.860} & 0.863 & 0.862 & - & - \\
4 & 72.11 & 75.60 & 77.43 & 77.97 & 77.02 & - & 0.716 & 0.752 & 0.772 & 0.777 & 0.767 & - & 0.815 & 0.844 & 0.860 & 0.861 & 0.850 & - \\
5 & 72.62 & 76.88 & \textbf{78.56} & \textbf{78.83} & 77.65 & 76.07 & 0.721 & 0.766 & \textbf{0.784} & \textbf{0.787} & 0.774 & 0.739 & 0.817 & 0.859 & \textbf{0.869} & \textbf{0.865} & 0.859 & 0.863\\
6 & 71.52 & 76.28 & 74.95 & 78.23 & 77.67 & 75.75 & 0.708 & 0.759 & 0.744 & 0.780 & 0.774 & 0.732 & 0.818 & 0.847 & 0.847 & 0.865 & 0.861 & 0.856\\
\hline
\end{tabular}
}
    \label{tab:ObservatioW}
\end{table*}

\begin{table}
\caption{\added[id=aby]{Confusion matrix of LSTM versus logistic regression using 
an observation window of $\tau=5$ sec and a buffer of $\eta=2$ sec.}}
\resizebox {\columnwidth} {!} {
    \begin{tabular}{c|| c c || c c }
\hline
& \multicolumn{2}{c||}{LSTM} & \multicolumn{2}{c}{Logistic Regression} \\
\hline
Classes & Engaged & SED & Engaged & SED \\
\hline
Engaged & 17745 & 2656 & 16479 & 3922 \\
SED & 6072 & 14329 & 5616 & 14785  \\
\hline
\end{tabular}
}
    \label{tab:ConfusionMatrix}
\end{table}

To better understand how performance is affected by the size $\tau$ of the observation window of the user behavior, we varied it \replaced[id=aby]{from 0 to 6 seconds}{in the range [1,6] seconds}. 
Table~\ref{tab:ObservatioW} shows this variation for each buffer $\eta$. 
For real-time operation using $\eta=0$, the best results were found using short observation windows of $\tau=1$ seconds for detecting SED. 
Increasing the buffer duration up to 3 seconds improves the performance of the SED detector. 
The best performance was found using an observation window of $\tau=5$ seconds for a buffer $\eta$ of 3 seconds and at approximately the same performance for a buffer of 2 seconds. We note that taking a buffer duration to make a decision approximately in the middle of the observation window is the best strategy to detect SED, and the optimal size of the observation window is inferior to the average duration of SED segments (\emph{i.e.} 6 seconds).  
\added[id=aby]{Table~\ref{tab:ConfusionMatrix} shows that logistic regression presents 30\% more false alerts than LSTM and 7\% fewer undetected engagement decreases.}


\section{Discussion}
\label{sec:discussion}
In order to develop lifelike humanoid robots that understand better the behavior of the humans with whom they interact and can respond appropriately to increase user engagement, 
we investigated the use of deep networks to successfully detect SED. We achieved good performance: 78\% of accuracy, 0.78 of F1-score and 0.87 of AUC. 
Note that in other related studies, the performances of engagement detection systems, using different datasets, were 62\% accuracy and 0.61 F1-score in \cite{Foster2017} and 73\% accuracy and 0.62 AUC in \cite{Leite2015}. 
\added[id=aby]{Thus we are using a bigger data-set with different annotation schema. But, we achieve promising results that could be improved and integrated in the robot architecture to detect SED with real-time capability.}

The classifiers provide not only the class of user engagement, but also the estimated confidence that could be used as additional information, representing the system's uncertainty, in real-world human-robot applications.

In preliminary experiments using less data (\emph{e.g.} 195 interactions), the best performances for GRU/LSTM using a buffer of $\eta=1$ second and an observation widow of $\tau=2$ seconds were 76\%/75\%, 0.75/0.75 and 0.84/0.84 for accuracy, F1-score and AUC, respectively. Thus, the GRU gives a slightly better performance. 

We evaluate the impact of two different extractors: OpenFace \cite{Baltrusaitis2018}, and Pepper's OKAO\textsuperscript{TM} Vision software \cite{OMRON} tracker of gaze direction, head motion and facial expression/AU. Table~\ref{tab:tracker} compares the performance of these extractors on the task of detecting SED using LSTM with an observation window of $\tau=5$ seconds and a buffer of $\eta=2$ seconds. We found that OpenFace performs better than Pepper's tracker. 
\added[id=aby]{Note that when features are missing (\textit{e.g.} when the robot's head is moving and user's facial features cannot be \replaced[id=aby]{determined}{extracted}), we focus on the other modality (i.e. distance, speech) to detect SED.}
\begin{table}[tb!]
\caption{Results using LSTM with an observation window $\tau=5$ seconds and a buffer of $\eta=2$ second with gaze direction, head motion, and facial expression extracted using OpenFace versus Pepper's OKAO\textsuperscript{TM} Vision software, combined with speech and distance streams.}
\resizebox {\columnwidth} {!} {
    \begin{tabular}{l c c c}
\hline
Tracker & Accuracy & F1-score & AUC \\
\hline
OpenFace \cite{Baltrusaitis2018} & 78.56 & 0.784 & 0.869 \\
Pepper OKAO software \cite{OMRON} & 76.33 & 0.762 & 0.849 \\
\hline
\end{tabular}
}
    \label{tab:tracker}
\end{table}

In spontaneous HRI, finding the exact moment of SED is a hard decision. It depends on the head motion, the looking away, the spoken word, getting away from the robot, etc. 
The annotated start and end of this segment is flexible and could vary by $\pm n$ frames (see Fig.~\ref{fig:interaction2min} in \nameref{sec:AppendixB}). 
It would be interesting to take into account this flexibility both in the training and in the testing phases instead of using it only when the annotators agree and ignoring the parts where they disagree.

Future work should also investigate whether the SED detection model generalizes well to other interaction settings (\emph{i.e.} other scenarios, multiparty). 

\section{Conclusion}
\label{sec:conclusion}


We analyzed users' behavior in two engagement states where they exhibited engaged behavior or, alternatively, signs of engagement decrease. We found significant differences in their behavior that allowed us to develop a real-time detector of engagement decrease during a spontaneous interaction with a humanoid robot. 

We then studied the use of deep learning techniques with multimodal data for real-time detection of user engagement decrease. 
Our engagement classification results show that the real-time detector taking into account the past user behavior without any buffer performs well. Using the temporal dynamics of user behavior improves the results as well. 
The optimal size of the observation window of user behavior is found to be smaller than the average duration of SED segments (\emph{i.e.} 6 seconds). 
Moreover, by using a delay of 1 or 2 seconds, we improved the performance of the detector.
Depending on the application context, these delays could be reasonably suitable to improve the experience quality of interacting with the robot in-the-wild. 

Finally, we believe that the publicly available dataset that we have collected \cite{Ben-Youssef2017}, presents a high potential for other tasks in human-robot interaction (\textit{e.g.} analysis of the social relationship between the user and the robot
).

\section{Appendix A}
\label{sec:AppendixA}


Survey of Satisfaction presented as the final phase of the scenario.
The participant was asked to indicate: 
\begin{enumerate}
\item his satisfaction with the interaction,
\item his involvement in the interaction,
\item his desire to leave the interaction, 
\item his desire to continue the interaction during the welcome phase,
\item his desire to continue the interaction during the dialog phase, 
\item his desire to continue the interaction during the cucumber phase, 
\item his desire to continue the interaction during the survey phase,
\item his desire to stay during the interaction,
\item if he believes that the robot wanted to stay during the interaction,
\item his desire to continue the conversation,
\item if he believes that the robot wanted to continue the conversation,
\item his feeling about his involvement in the interaction,
\item if he finds that the interaction was boring or fun,
\item  if he finds that the information was interesting,
\item if he liked the interaction.
\end{enumerate}

\section{Appendix B}
\label{sec:AppendixB}

\begin{acknowledgements}
This work was supported by European projects H2020 ANIMATAS (ITN 7659552) and a grant overseen by the French National Research Agency (ANR-17-MAOI). The authors would like to thank Nicolas Rollet and Christian Licoppe for useful discussions on pre-closing and Rodolphe Gelin, Angelica Lim, Marine Chanoux and Myriam Bilac from Softbank robotics for their help in the recording of UE-HRI dataset. 
\end{acknowledgements}

\section*{Compliance with Ethical Standards:}
\subsection*{Funding:}
This work was supported by SoftBank Robotics.

\subsection*{Conflict of interest:}
The authors declare that they have no conflict of interest.

\bibliographystyle{spmpsci}      

\bibliography{references}


\end{document}